\documentclass[preprint,12pt]{elsarticle}
\usepackage{amssymb}
\usepackage{graphicx}
\journal{Physics Letters B}
\begin{document}

\begin{frontmatter}

\title{Solving the relativistic rankine-hugoniot condition in presence of magnetic field in astrophysical scenario}

\author{Ritam Mallick}
\ead{ritam@physics.iisc.ernet.in}
\address{Department of Physics, Indian Institute of Science, 
        Bangalore 560012, INDIA}

\begin{abstract}
Rankine-Hugoniot condition has been solved to study phase transition in astrophysical scenario mainly in the case of phase transition from neutron
star (NS) to quark star (QS). The phase transition is brought about by a
combustion front, which travels from the center to the surface. 
The equations of state and temperature plays a huge
role in determining the nature of the front propagation, which brings about
the phase transition in neutron stars (NSs). Magnetic field has been introduced 
and the modified conservation condition for the perpendicular 
and oblique shocks is obtained. Numerical solution of the perpendicular 
shock has been 
shown in the figures, which finds that the magnetic field helps in shock
generation. It indirectly hints at the instability of the matter and thereby
the NS for very high magnetic field, implying that NSs can only support finite
magnetic field strength.
\end{abstract}

\begin{keyword}
stars: neutron, stars: magnetic fields, equation of state
\end{keyword}

\end{frontmatter}

\section{Introduction}
When the velocity of a fluid in motion becomes comparable with or exceeds 
that of the sound, the effect due to compressibility of the fluid become of 
prime importance. For a wave propagating in a non conducting gas, when 
the amplitude is 
so small that linear theory applies, the disturbance propagates as a sound 
wave. If the gas has a uniform pressure and density, the speed of propagation 
of sound and the wave profile maintains a fixed shape, since each part of 
the wave moves with same speed. However, when the wave possesses a finite 
amplitude, so that nonlinear terms in the equation become important, the 
crest of the sound wave moves faster than its leading or trailing edge. This 
causes a progressive steepening of the front portion of the wave as the 
crest catches up and ultimately, the gradient of pressure, density, 
temperature and velocity become large that dissipative processes, such as 
viscosity or thermal conduction are no longer negligible. Then a steady wave 
shape is attained, called a shock wave. 
The shock wave moves at a speed in 
excess of the sound speed, so the information cannot be propagated ahead to 
signal its imminent arrival, since such information would travel at sound 
speed relative to the undisturbed medium ahead of the shock. The dissipation 
inside the shock front leads to a gradual conversion of the energy being 
carried by the wave into heat. Thus, the effect of the passage of a shock 
wave are to convert ordered (flow) energy into random (thermal) energy 
through particle collisions and also to compress and heat the gas.
The shock front itself is in reality a very thin transition region. Its width 
is typically only a few mean-free paths, with particle collisions establishing 
the new uniform state behind the shock. 

The relativistic shock propagates into a medium with a changing equation of 
state. Therefore, a simple analysis of the jump condition for a polytropic 
or perfect fluid is not adequate and a deep understanding of this problem 
calls on for the full theoretical description of the relativistic shock in a 
medium with arbitrary equation of state. Further complication might 
arise if there is a presence of significant magnetic field. In fact, one 
can show that the relative importance of a magnetic field can grow during 
a collapse. In the field of nuclear physics, high energy collisions among 
heavy ions can be modeled by using fluid dynamical concepts. Also, some 
current models under investigation predict that relativistic shocks (or 
relativistic detonation and deflagration) might be related to the phase 
transition from nuclear matter to a quark matter.

Relativistic shock waves have been the subject of early investigation in 
relativistic fluid dynamics and magneto-fluid dynamics. In relativistic 
fluid dynamics the pioneering work is that of Taub \cite{taub} where the 
relativistic form of jump condition is established. A detailed analysis of 
the thermodynamic treatment of classical shock waves, is due to Thorne 
\cite{thorne}. Explicit solutions of the jump condition have ben obtained 
for special equation of states. Shock wave in relativistic magneto-fluid 
have been investigated extensively and in rigorous mathematical way by 
Lichnerowicz. \cite{lichnero}. Detonation and deflagration waves in 
relativistic magneto-fluid dynamics for nuclear physics and cosmology have been investigated by  \cite{stein,cleymans}. 

In Astrophysical scenario shock plays very important role in determining the outcome of compact stars.
In the case of massive stars, in the range between 8-100 solar masses, which 
are thought to be the progenitors of type II supernovas, one of the most 
viable mechanism for producing a explosion is gravitational collapse and 
bounce \cite{van}. In this case a shock is formed outside the inner 
core, which propagates outwards reaching relativistic speeds. Shock waves are 
also responsible for phase transition and gamma ray bursts (GRB) in 
compact stars.

In this paper I will mostly concentrate on the phase transition scenario in 
a compact star. Witten \cite{key-1} conjectured of 
strange quark matter (SQM), consisting of approximately equal numbers of 
up ({\it u}), down ({\it d}) and strange ({\it s}) quarks, is believed to be the  
ground state of strong interaction. This was  
supported by model calculations for certain ranges of values for strange 
quark mass and strong coupling constant \cite{key-2}. After that there has been 
constant efforts at confirming the existence of  
and SQM, though transiently, in ultra relativistic collisions. On the other 
hand, SQM could naturally occur in the cores of compact stars, 
where central densities are expected to be an order of magnitude higher 
than the 
nuclear matter saturation density. Thus, neutron stars which have  
sufficient high central densities might convert to strange star, 
or at least hybrid (a star with a quark core) stars. These transitions 
may lead to various observable 
signatures in the form of a jump in the breaking index and gamma ray 
bursts  \cite{key-3,key-4}, and a full QS might help in explaining 
the phenomena of observed quasi periodic oscillations \cite{key-4a}.

\par
There may be several scenarios by which neutron stars could convert
to quark stars. It may happen through a "seed" of external SQM \cite{key-5}, or 
triggered by the rise in the central density due to a sudden spin-down 
in older neutron stars \cite{key-6}. Several authors have studied the
conversion of nuclear matter to strange matter under different 
assumptions \cite{key-7,key-8,key-9,key-10,key-11,key-12,key-13,
key-14,key-15,key-16,key-17}. They have been summarized in a recent
work of ours \cite{key-18} and for the constraint of space, I do not
repeat them here.

After the discovery magnetars, some compact stars were found to have very
high surface magnetic fields. So the study of relativistic rankine-hugoniot 
condition is not sufficient. To have a full understanding of the properties
of NS and its phase transition to QS, such conditions should be examined in the presence of high magnetic fields. In this paper I wish to carry out such a 
basic calculation keeping our focus mainly on the astrophysical scenario.
The paper is organized as followed: first I will 
discuss the general rankine hugoniot condition as a discontinuity in the
conversion front. In section III, I will introduce magnetic field and study 
the new set 
of modified conservation equations. Then in section IV, I shall show my results and finally I will discuss and summarize them. 

\section{General Rankine-Hugoniot condition}

In this section I will discuss the  general rankine-hugoniot condition as the 
conservation equations which balances the conversion of neutron proton (n-p)
matter to two-flavour quark matter, consisting of u and d quarks along
with electrons for ensuring charge neutrality. I heuristically
assume the existence of a combustive phase transition front. Using the 
macroscopic
conservation conditions, I examine the range of densities for which
such a combustion front exists. 

Let us consider the physical situation where a combustion front
has been generated in the core of the Neutron star. This front propagates
outwards through the neutron star with a certain velocity, 
leaving behind a u-d-e matter. In the following, I denote all the physical
quantities in the hadronic sector by subscript 1 and those in the
quark sector by subscript 2. 
The conservation condition for energy-momentum and baryon number relates the
quantities on the opposite sides of the front.
In the rest frame of the
combustion front, these conservation conditions is given by
\cite{key-17,key-23,key-23a}:

\begin{equation}
\omega_{1}v_{1}^{2}\gamma_{1}^{2}+p_{1}=\omega_{2}v_{2}^{2}\gamma_{2}^{2}+p_{2},
\label{2}\end{equation}

\begin{equation}
\omega_{1}v_{1}\gamma_{1}^{2}=\omega_{2}v_{2}\gamma_{2}^{2},
\label{3}\end{equation}
and
\begin{equation}
n_{1}v_{1}\gamma_{1}=n_{2}v_{2}\gamma_{2}.
\label{4}\end{equation}

In the above three conditions $v_{i}$ (i=1, 2) is the velocity, $p_{i}$
is the pressure, $\gamma_{i}=\frac{1}{\sqrt{1-v_{i}^{2}}}$ is the
Lorentz factor, $\omega_{i}=\epsilon_{i}+p_{i}$ is the specific enthalpy
and $\epsilon_{i}$ is the energy density of the respective phases.

The velocities of the matter in the two phases, given by equations
(\ref{2}-\ref{4}), can be solved, such that \cite{key-23}:

\begin{equation}
v_{1}^{2}=\frac{(p_{2}-p_{1})(\epsilon_{2}+p_{1})}{(\epsilon_{2}
-\epsilon_{1})(\epsilon_{1}+p_{2})},
\label{5}\end{equation}

and \begin{equation}
v_{2}^{2}=\frac{(p_{2}-p_{1})(\epsilon_{1}+p_{2})}{(\epsilon_{2}
-\epsilon_{1})(\epsilon_{2}+p_{1})}.
\label{6}\end{equation}

It is possible to classify the various conversion mechanisms by comparing
the velocities of the respective phases with the corresponding velocities
of sound, denoted by $c_{si}$, in these phases. Thes conditions are
summarized in \cite{key-23c}.

For the conversion to be physically possible, velocities should satisfy
an additional condition, namely, $0\leq v_{i}^{2}\leq 1$. Here I
find that the velocity condition puts severe constraint on the 
allowed equations of state.

\section{Magnetic field Inclusion}
In a conducting gas, a magnetic field can interact strongly with the flow. The 
analysis of the shock waves therefore becomes more complex, but the basic principles remains the same. A set
of jump condition can again be derived, but they are considerably more complicated than the pure hydrodynamic shock case. The extra complexity arises both from the
presence of extra variable, namely the magnetic field strength, and also from the fact that the magnetic field and the matter velocities may be inclined with 
the shock normal.

A shock propagating through an magnetic fluid produces a significant difference in matter properties on either side of the shock front. The thickness of the front is determined by a balance between convective and dissipative effects. However, dissipative effects at high temperature are only comparable to convective effects when the spatial gradients in matter variables become extremely large. Hence, shocks in such matter tend to be extremely narrow, and are well-approximated as discontinuity. The hydrodynamical equations, and Maxwell's equations, can be integrated across a shock to give a set of jump conditions which relate matter properties on each side of the shock front. If the shock is sufficiently narrow then these relations become independent of its detailed structure.

In the rest frame of the shock, the shock front coincides with the $y$-$z$ plane. Furthermore, the regions of the plasma upstream and downstream of the shock, which are termed regions 1 and 2, respectively, be spatially uniform and non-time-varying. It follows that $\partial/\partial t = \partial/\partial y =\partial/\partial z=0$. Moreover, $\partial/\partial x=0$, except in the immediate vicinity of the shock. Finally, let the velocity and magnetic fields upstream and downstream of the shock all lie in the $x$-$y$ plane. The magnetic field is 
given by $B_i$ for the respective phases.

The first nontrivial shock is called perpendicular shock in which both the upstream and downstream plasma flows are perpendicular to the magnetic field, as well as the shock front. The conservation condition are given by 
 
\begin{equation}
\omega_{1}v_{1}^{2}\gamma_{1}^{2}+p_{1}+ \frac{{B_1}^2}{8\pi} =\omega_{2}v_{2}^{2}\gamma_{2}^{2}+p_{2}+\frac{{B_2}^2}{8\pi},
%\label{2}
\end{equation}

\begin{equation}
\omega_{1}v_{1}\gamma_{1}^{2}+v_1\gamma_1\frac{{B_1}^2}{4\pi}=\omega_{2}v_{2}\gamma_{2}^{2}+v_2\gamma_2\frac{{B_2}^2}{4\pi},
%\label{3}
\end{equation}

\begin{equation}
B_1v_1\gamma_1=B_2v_2\gamma_2
%\label{5}
\end{equation}

and 

\begin{equation}
n_{1}v_{1}\gamma_{1}=n_{2}v_{2}\gamma_{2}.
%\label{4}\
\end{equation}
 
The first three equation can be reduced to two equation, given by

\begin{equation}
\omega_{1}v_{1}^{2}\gamma_{1}^{2}+p_{1}+ \frac{{B_1}^2}{8\pi} =\omega_{2}v_{2}^{2}\gamma_{2}^{2}+p_{2}+\frac{{B_1}^2}{8\pi}(\frac{v_1\gamma_1}{v_2\gamma_2})^2,
%\label{2}
\end{equation}

\begin{equation}
\omega_{1}v_{1}\gamma_{1}^{2}+v_1\gamma_1\frac{{B_1}^2}{4\pi}=\omega_{2}v_{2}\gamma_{2}^{2}+v_2\gamma_2\frac{{B_1}^2}{4\pi}\frac{(v_1\gamma_1)^2}{v_2\gamma_2},
%\label{3}
\end{equation}

I now solve for $v_1$ nad $v_2$.

The most general shock is the oblique shock in which the plasma velocities and the magnetic fields on each side of the shock are neither parallel nor perpendicular to the shock front. The Rankine-Hugoniot condition are given by 

\begin{equation}
\omega_{1}v_{1x}^{2}\gamma_{1}^{2}+p_{1}+ \frac{{B_1}^2}{8\pi}-
\frac{{B_{1x}}^2}{4\pi} = \omega_{2}v_{2x}^{2}\gamma_{2}^{2}+  
p_{2}+\frac{{B_2}^2}{8\pi}-\frac{{B_{2x}}^2}{4\pi},
%\label{2}
\end{equation}

\begin{equation}
\omega_{1}v_{1x}v_{1y}\gamma_{1}^{2}+v_{1x}\gamma_1\frac{B_{1x}B_{1y}}{4\pi}=\omega_{2}v_{2x}v_{2y}\gamma_{2}^{2}+v_{2x}\gamma_2\frac{B_{2x}B_{2y}}{4\pi},
%\label{3}
\end{equation}

\begin{equation}
\omega_{1}v_{1x}\gamma_{1}^{2}+v_{1x}\gamma_1\frac{{B_1}^2}{4\pi}-\frac{B_{1x}(B_1.v_1)\gamma_1}{4\pi}=\omega_{2}v_{2x}\gamma_{2}^{2}+v_{2x}\gamma_2\frac{{B_2}^2}{4\pi}-\frac{B_{2x}(B_2.v_2)\gamma_2}{4\pi},
%\label{3}
\end{equation}

\begin{equation}
B_{1y}v_{1x}\gamma_1-v_{1y}B_{1x}\gamma_1=B_{2y}v_{2x}\gamma_2-v_{2y}B_{2x}\gamma_2
%\label{5}
\end{equation}

\begin{equation}
B_{1x}=B_{2x}
\end{equation}

and 

\begin{equation}
n_{1}v_{1x}\gamma_{1}=n_{2}v_{2x}\gamma_{2}.
%\label{4}
\end{equation}
 
There may be two easier cases for the above complicate equation.

Case 1. $v_{1y}=0=v_{2y}$

Then the first four equation would simplify to 

\begin{equation}
\omega_{1}v_{1x}^{2}\gamma_{1}^{2}+p_{1}+ \frac{{B_1}^2}{8\pi} =\omega_{2}v_{2x}^{2}\gamma_{2}^{2}+p_{2}+\frac{{B_2}^2}{8\pi},
%\label{2}
\end{equation}

\begin{equation}
\omega_{1}v_{1}\gamma_{1}^{2}+v_{1}\gamma_1\frac{{B_1}^2}{8\pi}-\frac{B_{1x}(B_1.v_1)\gamma_1}{4\pi}=\omega_{2}v_{2}\gamma_{2}^{2}+ 
v_{2}\gamma_2\frac{{B_2}^2}{8\pi}-\frac{B_{1x}(B_2.v_2)\gamma_2}{4\pi},
%\label{3}
\end{equation}

\begin{equation}
B_{2y}=\frac{v_1\gamma_1}{v_2\gamma_2}B_{1y}
\end{equation}

Case 2.  $B_{1y}=0=B_{2y}$

\begin{equation}
\omega_{1}v_{1x}^{2}\gamma_{1}^{2}+p_{1} =\omega_{2}v_{2x}^{2}\gamma_{2}^{2}+p_{2},
%\label{2}
\end{equation}

\begin{equation}
\omega_{1}v_{1x}\gamma_{1}^{2}+v_{1x}\gamma_1\frac{{B_1}^2}{8\pi}-\frac{B_{1x}(B_1.v_1)\gamma_1}{4\pi}=\omega_{2}v_{2x}\gamma_{2}^{2}+ 
v_{2x}\gamma_2\frac{{B_2}^2}{8\pi}-\frac{B_{1x}(B_2.v_2)\gamma_2}{4\pi},
%\label{3}
\end{equation}

\begin{equation}
\omega_1v_{1x}v_{1y}{\gamma_1}^2=\omega_2v_{2x}v_{2y}{\gamma_2}^2
\end{equation}

\section{Results}

I start my calculations by using the nuclear matter EOS obtained through
nonlinear Walecka model \cite{wal}.
In the present paper, I consider the conversion of nuclear
matter, consisting of only nucleons, 
to a two-flavour quark matter. The final composition of the quark matter
is determined
from the nuclear matter EOS by enforcing the baryon number conservation
during the conversion process.
While describing the state of matter for the
quark phase I consider a range of values for the bag constant. Nuclear
matter EOS is calculated at zero temperature, whereas, the two-flavour
quark matter EOS is obtained both at zero temperature as well as at
finite temperatures as during the propagation of the shock it may heat up 
the matter.

To examine the nature of the hydrodynamical front, arising from the
neutron to two-flavour quark matter conversion, I plot, in fig.1,
the quantities $v_{1},v_{2},c_{s1}$ and $c_{s2}$ as functions of
the baryon number density ($n_{B}$). As mentioned earlier, the u
and d quark content in the quark phase is kept same as the one corresponding
to the quark content of the nucleons in the hadronic phase. With these
fixed densities of the massless u and d quarks and electrons, the EOS of the 
two-flavour 
matter has been evaluated using the bag model prescription.
I find that the velocity condition ( ${v_{i}}^{2}>0$ ) is satisfied only 
for a small
window of $\approx\pm5.0MeV$ around the bag pressure $B^{1/4}=160MeV$.
The constraint imposed by the above conditions results in the possibility
of deflagration, detonation or supersonic front as shown in the 
fig. 1.

In fig. 1, I considered both the phases to be at zero temperature.
A possibility, however, exist that a part of the internal energy is
converted to heat energy, thereby increasing the temperature of the
two-flavour quark matter during the exothermic combustive conversion
process.  In
fig. 2, I plot the variation of velocities with density
at $T=50MeV$, for which significant change is noticed. This figure show 
that the range of values of baryon
density, for which the flow velocities are physical, increases with
temperature. In the present paper I have considered only
the zero temperature nuclear matter EOS. 
On the other hand, equation of state of quark matter
has a finite temperature dependence and hence the difference between
${v_{1}}$ and ${v_{2}}$, varies with temperature.

The variation of the velocities with temperature is due to the fact that
higher temperature means higher energy. As the energy increases the 
particles becomes more energetic which means the matter becomes more 
excited and compressible. 
As matter becomes more compressible due to increase in temperature, now there 
is a chance for shock formation which was previously not possible. 
  
\begin{figure}
\vskip 0.3in
   \centering
\includegraphics[width=3.0in]{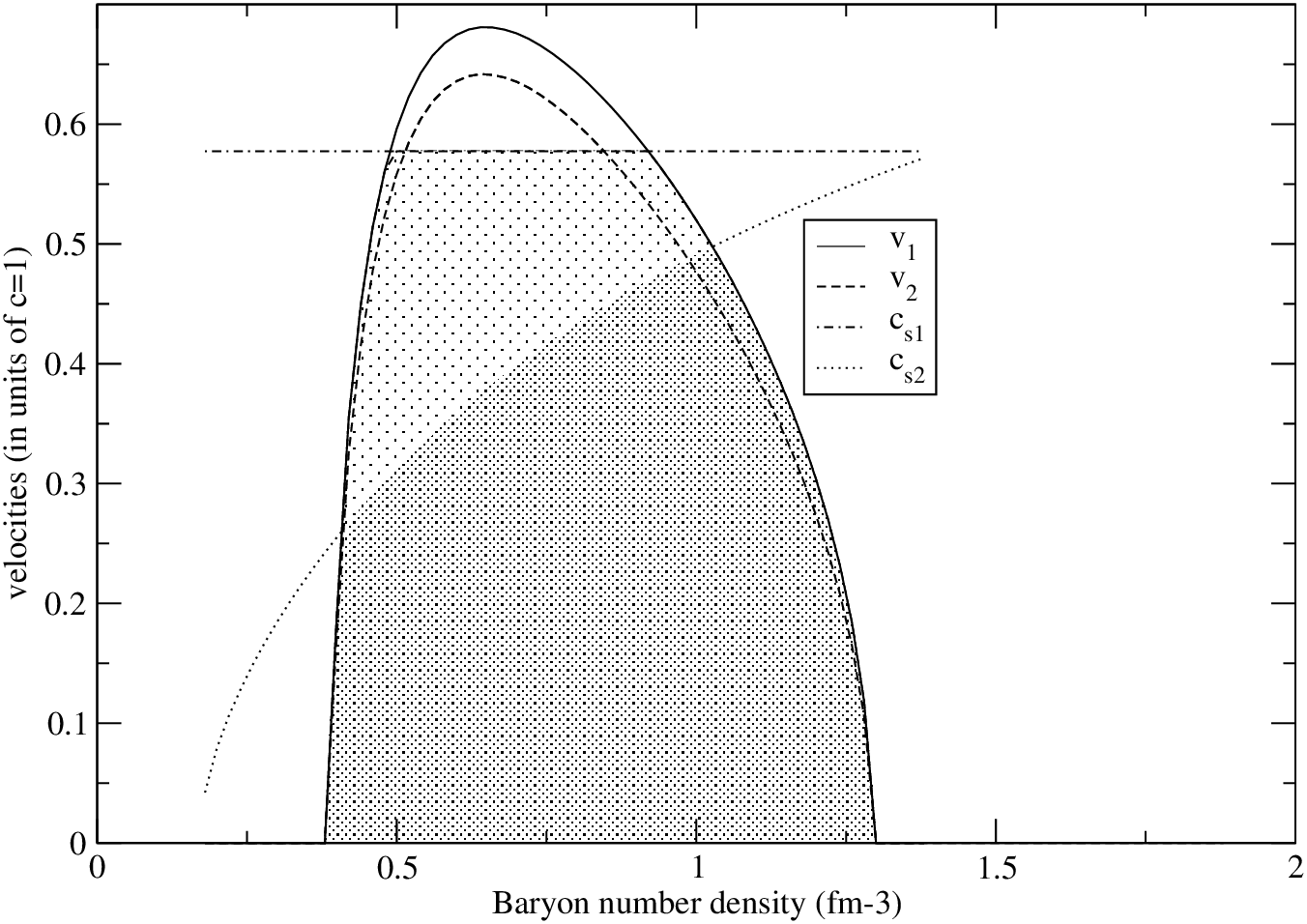}
\caption{Variation of different velocities with baryon number density for
T = 0 MeV, $B^{1/4}=160 MeV$ and strange quark mass $m_s=200 MeV$. 
The dark-shaded region correspond to 
deflagration, light-shaded
region correspond to detonation and the unshaded region
correspond to supersonic conversion processes.}
\end{figure}

\begin{figure}
\vskip 0.3in
   \centering
\includegraphics[width=3.0in]{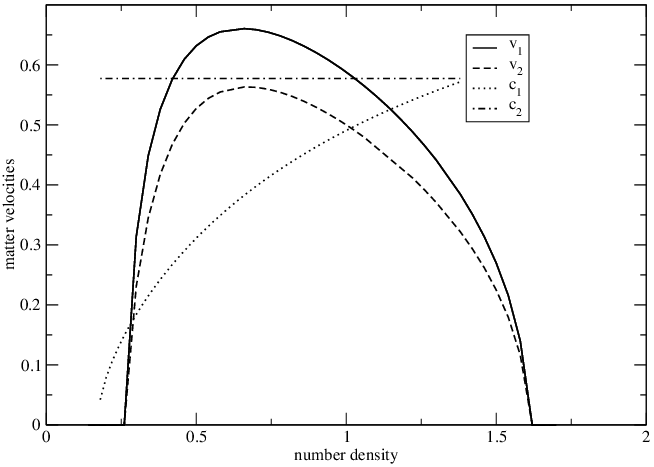}
\caption{Variation of velocities with baryon number density for T = 80 MeV
 $B^{1/4}=160 MeV$ and strange quark mass $m_s=200 MeV$.
Different regions correspond to different modes of conversion similar to that of
fig 1.}
\end{figure}

Till now I have used equation for the normal hydrodynamics for the generation
of the curves. The change in temperature in the EOS accounted for the change in the nature of the graph. Now I will plot graphs solving magnetic field 
induced hydrodynamics. I solve the new conservation conditions for the 
perpendicular shock in which both the upstream and downstream plasma flows are perpendicular to the magnetic field, as well as the shock front. 

In fig. 3 I have plotted curves for different velocities with baryon density
for the zero temperature case. The magnetic field used for the generation of 
the curve is $B=5\times10^{15}G$.  I find that due to introduction of the 
magnetic field the range of values of baryon
density, for which the flow velocities are physical, increases. Its nature is
quite similar to that for the case of temperature. Due to the introduction of 
the magnetic field the the pressure increases (pressure due to magnetic field is
$B^2/8\pi$) for the same
value of baryon number density. The magnetic field in the respective phases 
adjust in such a way that the resultant energy and pressure of the both 
phases gets modified to make the range of baryon number density to increase. 
For both higher and lower values of baryon density,
previously there was less chance of shock formation, but now due to the
introduction of the magnetic field there is a greater chance of shock formation.
And therefore the range of baryon density gets much wider.    
 
Next I plot curves for different velocities for the finite temperature case
$T=80 MeV$. For same value of the magnetic field 
$B=5\times10^{15}G$, the range of baryon density gets much more wider. This 
is due to the fact that now both the temperature and the magnetic field 
work hand by hand to increase the chances of shock generation. 
Both process ensures by its own way that the matter parameters 
adjust itself in such a way that there is a greater probability of 
shock generation.

In fig. 5 I have plotted exclusively $v_1$ and $v_2$ for two cases, one without magnetic field and the other with magnetic field. The nature of the curve remains same, that is $v_1$ is always greater than $v_2$, which means the shock front propagates outward of the star. 
The range of baryon density, for which the matter velocities are finite, 
increases with magnetic field.
I have plotted this for the zero temperature case.
In fig. 6 I have plotted the same for the finite temperature case and find that for lesser value of magnetic field, same increase in range of baryon number density is seen. All the reasons for this nature is explained in the previous paragraph.

In fig 7. I have plotted $v_1$ for different values of magnetic field. I find that as the value of the magnetic field increases the range of baryon number density increases which is what was expected (and discussed previously). But as we go on increasing the magnetic field $v_1$ does not comes down on the lower
half of the curve. This is due to the fact that, at such high value of the magnetic field the matter becomes unstable. At other half where matter is at 
much higher density, it can support such field strength. But, if we 
further increase the magnetic field the matter cannot support such high 
fields whatever the density might be. The maximum value of magnetic field 
that matter can support is few times $10^{17}G$ (in our case the cut off 
value is $2\times 10^{17} G$). So I find that there is a cut off value for 
the magnetic field, and from here we can indirectly say that NS also can 
support up to a finite value of magnetic field. This curve is plotted for 
zero temperature, and in fig. 8 I have plotted the same for finite 
temperature. Qualitatively the nature of the graph remains the same only 
the quantitative value of the magnetic field changes. It shows that 
hotter matter can support lesser value of magnetic field than the colder one.   

\begin{figure}
\vskip 0.3in
   \centering
\includegraphics[width=3.0in]{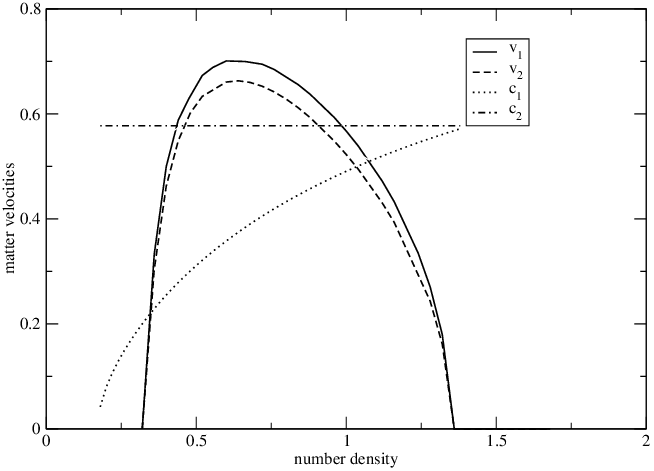}
\caption{Variation of different flow velocities with baryon number density 
for T = 0 MeV and magnetic field of
$B=5\times10^{15}G$.}
\end{figure}

\begin{figure}
\vskip 0.3in
   \centering
\includegraphics[width=3.0in]{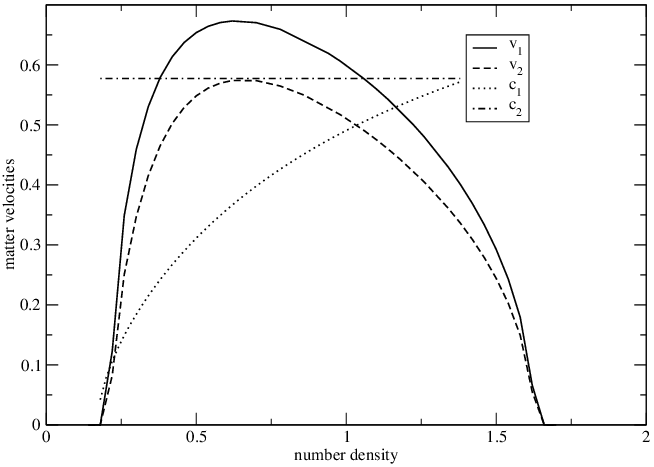}
\caption{Variation of velocities with baryon number density for $T = 80 MeV$, and magnetic field of $B=5\times10^{15}G$.}
\end{figure}

\begin{figure}
\vskip 0.3in
   \centering
\includegraphics[width=3.0in]{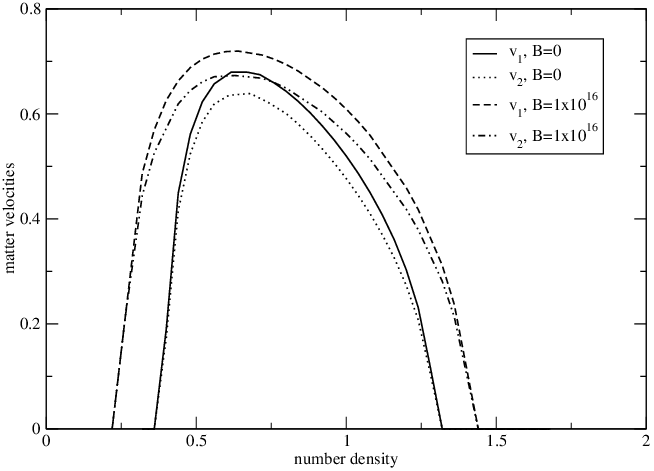}
\caption{Comparison of $v_n$ and $v_s$ with baryon number density for $T = 0 MeV$
 for two values of magnetic field strength $B=0G$ and $B=1\times10^{16}G$.}
\end{figure}

\begin{figure}
\vskip 0.3in
   \centering
\includegraphics[width=3.0in]{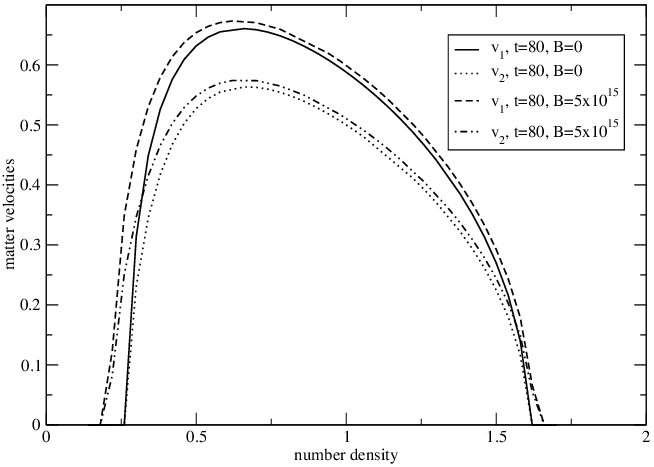}
\caption{Comparison of $v_n$ and $v_s$ with baryon number density for $T = 80 MeV$
 for two values of magnetic field strength $B=0G$ and $B=5\times10^{15}G$.}
\end{figure}

\begin{figure}
\vskip 0.3in
   \centering
\includegraphics[width=3.0in]{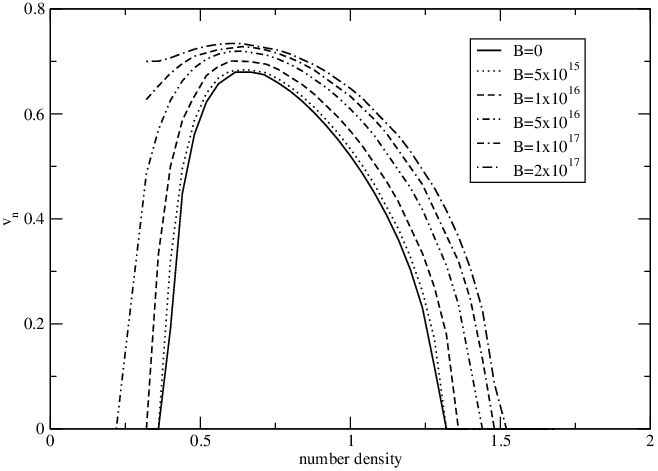}
\caption{Comparison of $v_n$ with baryon number density for $T = 0 MeV$
 for different values of magnetic field strength.}
\end{figure}

\begin{figure}
\vskip 0.3in
   \centering
\includegraphics[width=3.0in]{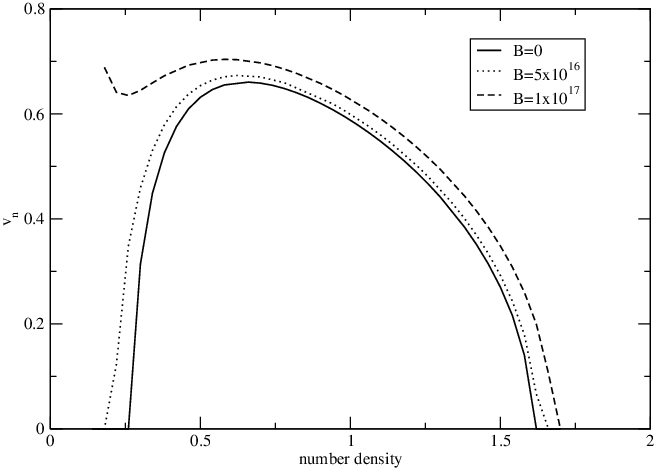}
\caption{Comparison of $v_n$ with baryon number density for $T = 80 MeV$
 for different values of magnetic field strength.}
\end{figure}

\section{Summary and discussion}

Finally in this section I summarize my results. I find that rankine-hugoniot condition can be solved to determine the condition for different types of wave generation in a neutron star. It also determines the mode of the propagation of the wave front. The temperature (finite temperature of the matter) helps in the generation of the front and as the temperature rises the wave front can generate both at much lower and at much higher baryon densities which was not possible for the zero temperature case. Next I write down the modified conservation conditions in presence of magnetic field. I have written down conditions for both the perpendicular and oblique shock waves. The inclusion of the magnetic field introduces not only extra conditions but also the earlier existing conditions gets modified. I have solved the velocity of the matter of the two phases and plotted curves for the perpendicular wave, which was not obtained analytically. The conditions cannot be solved analytically and and therefore I have solved the nonlinear simultaneously equation numerically. I have matched my nonmagnetic numerical 
results with the analytically solvable nonmagnetic solutions. The oblique wave equation gets very much complicated and the simultaneous equations does not converges for different value of the baryon densities. So I have not plotted the results of the oblique waves. 

Solving the perpendicular wave for finite magnetic field I find that the range of baryon density, for which the flow velocities of matter are physical, increases with magnetic field strength. This is because by the introduction of the magnetic field the resultant pressure and energy redistribute in such a way that, for same baryon density, the changes for shock generation increases. 
I also find that there is a cut off magnetic field strength that matter can support, beyond which the matter becomes unstable and the flow velocities becomes imaginary. This on the other hand suggest that a NS cannot support magnetic field beyond a certain field strength. To finally summarize our result I mention that this is the first instance where such a treatment of modified conservation
condition has been done in the presence of magnetic field in the astrophysical phase transition scenario. This provides with new interesting results, that has not been anticipated before and also indirectly hints at the instability of the NS at very high magnetic field. More interesting results is anticipated if the full oblique wave equations can be solved and I am now focussing mainly on that problem.

I would like to thank Grant No. SR/S2HEP12/2007, funded by DST, India for financial support.


\begin{thebibliography}{99}
\bibitem{taub} A. H. Taub, Physical Review, 74, 328 (1948)
\bibitem{thorne} K. S. Thorne, Astrophys. J., 179, 897 (1973)
\bibitem{lichnero} A. Lichnerowicz, 1967, {\it{Relativistic hydrodynamics and Magnetohydrodynamics}}, New York, Benjamin
\bibitem{stein} P. J. Steinhardt, Phys. Rev. D, 25, 2074 (1982)
\bibitem{cleymans} J. Cleymans, R. V. Gavai, E Suhonen, Physics reports, 130, 217 (1986)
\bibitem{van} K. A. Van Riper, Astrophys. J., 232, 558 (1979)
\bibitem{key-1} E. Witten, {\it Phys. Rev.} {\bf{D30}}, 272 (1984)
\bibitem{key-2} E. Farhi and R. L. Jaffe, {\it Phys. Rev.} {\bf{D30}}, 2379 
(1984)
\bibitem{key-3} A. Bhattacharyya, S. K. Ghosh, M. Hanauske and S. Raha, 
{\it Phys. Rev.} {\bf{C71}}, 048801 (2005)
\bibitem{key-4} A. Bhattacharyya,  S. K. Ghosh and S. Raha, {\it Phys. Lett.} 
{\bf{B635}}, 195 (2006)
\bibitem{key-4a} A. Bhattacharyya and S. K. Ghosh, {\it Mod. Phys. Lett.} 
{\bf{A22}}, 1019 (2007)
\bibitem{key-5} C. Alcock, E. Farhi and A. Olinto, {\it Astrophys. J.} 
{\bf{310}}, 261 (1986)
\bibitem{key-6} N. K. Glendenning, {\it Nucl. Phys. (Proc. Suppl.)} {\bf{B24}}, 
110 (1991);  {\it Phys. Rev.} {\bf{D46}}, 1274 (1992)
\bibitem{key-7} A. Olinto, {\it Phys. Lett.} {\bf{B192}}, 71 (1987); 
{\it Nucl. Phys. (Proc. Suppl.)} {\bf{B24}}, 103 (1991)
\bibitem{key-8} M. L.Olesen and J. Madsen, {\it Nucl. Phys. (Proc. Suppl.)} 
{\bf{B24}}, 170 (1991)
\bibitem{key-9} H. Heiselberg, G. Baym  and C. J. Pethick, {\it Nucl. Phys. 
(Proc. Suppl.)} {\bf{B24}}, 144 (1991)
\bibitem{key-10} G. Lugones, O. G. Benvenuto and H. Vucetich,  {\it Phys. Rev.} 
{\bf{D50}}, 6100 (1994)
\bibitem{key-11} O. G. Benvenuto, and J. E. Horvarth, {\it Phys. Lett.} 
{\bf{B213}}, 516 (1988)
\bibitem{key-12} O. G. Benvenuto, J. E. Horvarth and H. Vucetich, 
{\it Int. J. Mod. Phys.} {\bf{A4}}, 257 (1989); O. G. Benvenuto and J. E. 
Horvarth, {\it Phys. Rev. Lett.} {\bf{63}}, 716 (1989) 
\bibitem{key-13} H. T. Cho, K. W. Ng and A. W. Speliotopoulos, {\it Phys. Lett.} 
{\bf{B326}}, 111 (1994)
\bibitem{key-14} I. Tokareva, A. Nusser, V. Gurovich and V. Folomeev, {\it Int. J. 
Mod. Phys.} {\bf{D14}}, 33 (2005) 
\bibitem{key-15} Z. Berezhiani, I. Bombaci, A. Drago, F. Frontera and A. Lavagno, 
{\it Astrophys. J.}  {\bf{586}}, 1250 (2003) 
\bibitem{key-16} I. Bombaci, I. Parenti and I. Vidana, {\it Astrophys. J.} 
{\bf{614}}, 314 (2004)
\bibitem{key-17} Tokareva, I., Nusser, A.,
Gurovich, V., Folomeev, V., 2005, Int. J. Mod. Phys., D14, 33
\bibitem{key-18} A. Bhattacharyya, S. K. Ghosh, P. Joarder, R. Mallick and 
S. Raha, {\it Phys. Rev.} {\bf{C74}}, 065804 (2006)
\bibitem{key-23} Landau, L. D., and Lifshitz, E. M.,
1987, {\it{Fluid Mechanics}}, Pergamon Press, New York 
\bibitem{key-23a} Anile, A. M., 1989, 
{\it{Relativistic fluids and Magneto-fluids : with application
in Astrophysics and Plasma Physics}}, Cambridge University Press,
U.K. 
\bibitem{key-23c} Laine, M., 1994, Phys. Rev., D49, 3847 
\bibitem{wal} J. Ellis, J. I. Kapusta
and K. A. Olive,  Nucl. Phys. B, {\bf{348}}, 345 (1991)
\end{thebibliography}
\end{document}